%% file: main_arXiv.tex
\documentclass[%
twocolumn,
 amsmath,amssymb,
 aps,
pra,
longbibliography,
]{revtex4-2}

\usepackage{graphicx}
\usepackage{dcolumn}
\usepackage{comment}
\usepackage{xcolor}
\usepackage[colorinlistoftodos]{todonotes}
\usepackage[colorlinks=true,citecolor=blue,urlcolor=blue]{hyperref}
\usepackage[english]{babel}
\usepackage{bm}
\usepackage{amsmath}
\usepackage{balance}
\usepackage{flushend}
\usepackage{amssymb}
\usepackage{comment}
\usepackage{subfigure}
\usepackage{epstopdf}
\usepackage{appendix}
\usepackage{float}
\usepackage[normalem]{ulem}
\usepackage{mathtools}
\usepackage{physics}
\hypersetup{hidelinks}

\DeclareUnicodeCharacter{03B2}{\ensuremath{\beta}}
\UseRawInputEncoding

\begin{document}

\title{Efimov Effect in Ultracold Microwave-Shielded Polar Molecules}

\author{Shayamal Singh$^{1}$}
\author{Chris~H.~Greene$^{1,2}$}
\email{chgreene@purdue.edu}

\affiliation{$^{1}$Department of Physics and Astronomy, Purdue University, West Lafayette, Indiana 47907, USA }
\affiliation{$^{2}$Purdue Quantum Science and Engineering Institute, Purdue University, West Lafayette, Indiana 47907 USA}

\date{\today}

\begin{abstract}
A quantum-mechanical description is presented for the three-body physics of shielded dipolar molecules, including a prediction of observable Efimov physics. Despite the anisotropic and long-range nature of the interaction, shielding enables a regime in which universality emerges already at the two-body level and extends to the three-body sector, where Efimov physics emerges. On the negative side of the scattering-length resonance, computed trimer binding energies display the characteristic scaling expected for Efimov states. Finally, the sudden approximation can be used to create trimer bound states, starting from positive energy trap states as a way to create or detect these molecular trimers. Moreover, the three-body parameter expressed in dipolar units is found to be universal.
\end{abstract}

\maketitle
Ultracold polar molecules hold great promise as a flexible platform for various quantum science applications. However, progress toward these goals has been severely limited, since both reactive and non-reactive polar molecules typically undergo rapid inelastic loss at short range~\cite{ospelkaus2010quantum,bause2023ultracold}, preventing evaporative cooling. Recent advances in collisional shielding have fundamentally changed this landscape.

Shielding techniques~\cite{karman2018microwave,karman2025double,gorshkov2008suppression,cooper2009stable,lassabliere2018controlling,avdeenkov2006suppression,quemener2016shielding} use external electromagnetic fields to engineer repulsive barriers at long range, preventing molecules from accessing the lossy short-range region. Microwave shielding with a circularly polarized field~\cite{karman2018microwave,lassabliere2018controlling} has been shown to strongly suppress two-body losses~\cite{anderegg2021observation, bigagli2023collisionally, lin2023microwave,schindewolf2022evaporation}, enabling cooling to Fermi degeneracy~\cite{schindewolf2022evaporation}. For bosonic molecules, however, the long-range interaction has an S-wave component, which can be sufficiently attractive  to support two-body bound states, referred to as field-linked states~\cite{avdeenkov2003linking,lassabliere2018controlling}; in this regime, three-body recombination into these states dominates the loss dynamics~\cite{stevenson2024three}.
A decisive breakthrough was the development of double microwave shielding, which combines circularly and linearly polarized microwave fields to engineer an interaction that eliminates field-linked bound states, which has enabled the cooling of bosonic molecular gases to the critical temperature and creating the first dipolar Bose-Einstein condensates (BECs)~\cite{bigagli2023observation,shi2025bose}. As a result, molecular systems have entered a new regime in which collisional losses are strongly suppressed and few-body interactions can be explored in a controlled and tunable manner.\\
This progress raises a natural and fundamental question: what few-body quantum phenomena emerge in shielded dipolar molecular systems? In particular, by varying the microwave detuning, Rabi frequency, and ellipticity (and, in double shielding, the relative strength of the field components), the effective two-body interaction can be tuned, allowing control of the scattering length analogous in spirit to magnetic Feshbach tuning in atomic gases~\cite{chin2010feshbach}. In this regime, universal few-body physics, most notably the Efimov effect~\cite{efimov1970energy} is expected to arise. The Efimov effect occurs when three particles interact via short-range, nearly resonant attractive interactions, leading to an effective three-body attraction that supports an infinite series of weakly bound trimer states with a universal geometric energy scaling given by $E_{n+1}/E_n=e^{-2\pi/s_0}$, where for three-identical bosons the universal constant $s_0=1.00624$.

While Efimov physics has been extensively studied in atomic gases~\cite{BRAATENReview,DIncaoReview,NaidonReview,greene2017universal},  its realization in molecular systems remains largely unexplored due to the aforementioned experimental challenges. Previous theoretical studies of dipolar few-body physics have considered Efimov states and other universal bound states in non-shielded systems where anisotropic long-range dipolar interactions are combined with an isotropic short-range core~\cite{greene2017universal,oi2024universality,wang2011efimov}.

Microwave-shielded interactions differ qualitatively from these models, as the shielding core itself is repulsive and anisotropic. The resulting barrier details can be tuned through external field parameters such as microwave intensity, detuning, and ellipticity while the long-range tail is characterized by the dipolar length $a_d$, which is defined in terms of the effective dipole moment $d_{\text{eff}}$~\cite{lahaye2009physics, chomaz2023dipolar,schindewolf2025few}.
The anisotropic nature of both the dipole-dipole interaction and the shielding core, together with the presence of an external field, renders the three-body problem for shielded molecules particularly challenging, as the total angular momentum $J$ is no longer conserved. 

Recent studies of the shielded dipolar three-body problem include a classical trajectory treatment~\cite{stevenson2024three}, a quantum approach based on an effective one-dimensional (1D) model~\cite{shi2026universal,wang2025interaction}, and a shielding potential based approach in two-dimensions~\cite{huang2012field}. A key parameter for the classical description is the dipolar energy $E_d=1/m a_d^2$ in atomic units, where $m$ is the molecular mass. For heavy molecules like NaCs and at typical experimental temperatures, a large number of partial waves start contributing to the cross-section, which justifies the classical treatment, but it is expected to lose validity as the temperature approaches $k_B T \sim E_d$. The effective 1D model, in turn, is for single field shielding and non-zero ellipticity and, notably, predicts a suppression of the Efimov effect. By contrast, current shielding experiments typically employ nearly circular polarization (small ellipticity), which is favorable for suppressing two-body inelastic loss~\cite{karman2019microwave}, and to date they focus on a gas with a single molecular species.

This letter develops a quantum-mechanical treatment of three microwave-shielded molecules that incorporates experimentally realized microwave-dressed interactions in three dimensions and explicitly accounts for their anisotropic character. When simultaneous dressing by a linearly polarized $\pi$ field and a circularly polarized $\sigma_+$ field is considered, universality emerges already at the two-body level, as predicted in Ref.~\cite{karman2025double,dutta2025universality}. Explicitly, when the lengths and energies are scaled by $|a_d|$ and $E_d$, the two-body adiabatic potentials depend only on the dimensionless ratio $a_s/|a_d|$, while the variation of this ratio is found to be independent of the molecule, provided that it can be shielded. Extending this to the three-body system using the adiabatic hyperspherical framework, irrespective of the molecule, the three-body curves also only depend on $a_s/|a_d|$. Moreover, the Efimov effect emerges when the scattering length $a_s$ becomes the dominant length scale in the problem, with the trimer spectrum and position of Efimov resonances determined solely by $a_s/|a_d|$. Here, two-body universality refers to species-independent low-energy scattering in dipolar units, whereas three-body universality refers to the Efimov trimer spectrum in the $|a_s|\gg|a_d|$ regime. Finally, the formation probability for trimolecular bound state is estimated under experimentally realistic conditions using the sudden approximation. These results establish microwave-shielded dipolar molecules as a highly tunable platform for exploring few-body physics in molecular systems.

\begin{figure}[t]
    \includegraphics[width=8.6cm]{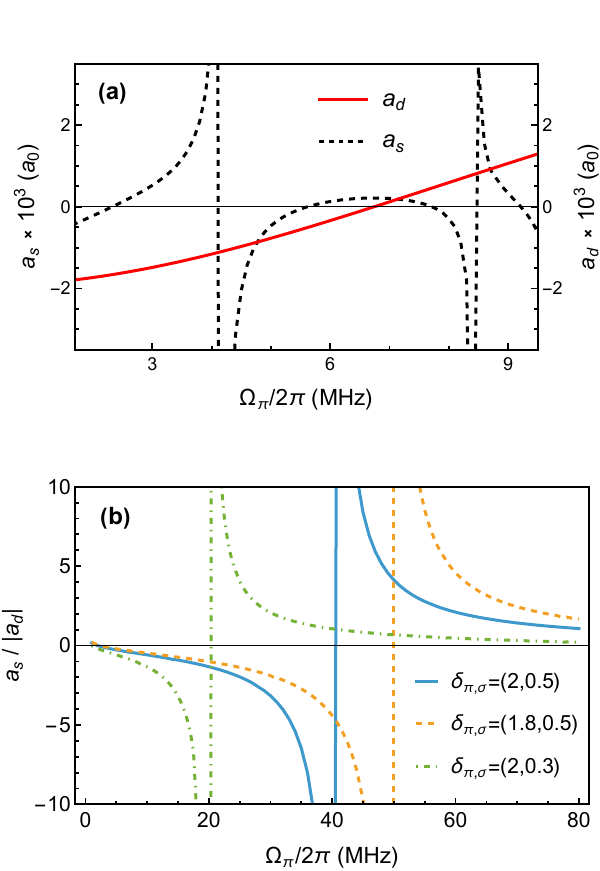}\\
    \caption{ (a) Typical variation of the scattering length $a_s$ (dashed) and dipolar length $a_d$ (solid) in scheme $\text{I}$ for NaCs as function of $\Omega_\pi$ with microwave field parameters $(\Omega_\sigma,\Delta_\sigma,\Delta_\pi)=2\pi\times(8,8,10)$ MHz . $a_d$ goes to zero at the compensation point where the $\sigma$ and $\pi$ field balance each other. On either side of the compensation point, there are field linked resonances. Increasing (Decreasing) $\Omega_\pi$ leads to $\pi$-~($\sigma$)-dominated  resonances. (b) Variation of $a_s/|a_d|$ for CaF in scheme $\text{II}$ for different $\delta_{\pi,\sigma}$ values with $\mathcal{G}=2$ and $a_d=-(3840.6,~3639.4,~4900)a_0$ for $\delta_{\pi,\sigma}=(2,0.5)$, $(1.8,0.5)$ and $(2,0.3)$ respectively.}
    \label{fig:1}
\end{figure}


  The effective interaction between two shielded molecules is obtained by projecting the dipole-dipole interaction onto the highest field-dressed eigenstate~\cite{deng2023effective}. For molecules dressed simultaneously by a circularly polarized $(\sigma^+)$ and a linearly polarized $(\pi)$ microwave field, the resulting effective potential $V_\text{eff}$ can be written using perturbation theory as~\cite{deng2025two}
\begin{multline}
    V_{\text{eff}}(\mathbf{r})=\frac{C_{3}}{r^3}(1-3\cos^2{\theta})+\frac{C_6}{r^6}\left(w_0(3\cos^2{\theta}-1)^2\right.\\ \left.+w_1\sin^2{\theta}\cos^2{\theta}+w_2\cos^4{\theta}\right),
\end{multline}

where $C_3, C_6$ and $w_i$ for $i=0,1,2$ depend on the Rabi frequency $\Omega_{\sigma,\pi}=(\Omega_\sigma,\Omega_\pi)$ and the detuning $\Delta_{\sigma,\pi}=(\Delta_\sigma,\Delta_\pi)$ of the microwave field, $r$ represents the intermolecular distance, $\hat{z}$ is the propagation direction of the $\sigma^+$-polarized microwave field, and the polarization direction of the $\pi$-polarized microwave field, and $\theta$ is the angle between the intermolecular axis and the $\hat{z}$ direction. In atomic units, $C_3= -d_0^2 / 12(1 + \delta_\sigma^2)$ for dressing with only the $\sigma_+$-field , and $C_3= d_0^2 / 6(1 + \delta_\pi^2)$ for dressing with only the $\pi$-field, where $d_0$ is the permanent dipole moment of the molecule, and $\delta_{\pi,\sigma}=(\Delta_\pi/\Omega_\pi, \Delta_\sigma/\Omega_\sigma)$. Qualitatively, this difference in the sign of the interaction is responsible for shaping the long range barrier in the case of simultaneous dressing. The dipolar length in atomic units is defined as $a_d=\mu_{2B}C_3$, such that $a_d<0~(a_d>0)$ corresponds to the $\sigma_+~(\pi)$ dominated field. The two-body coupled-channel equations are solved with this effective potential $V_\text{eff}$, using the slow-variable discretization (SVD) method~\cite{tolstikhin1996slow} along with the discrete variable representation (DVR) to propagate the R-matrix~\cite{wang2012hyperspherical,manolopoulos1988quantum}, and obtain the scattering length $a_s$~\cite{SI}. Since $V_\text{eff}$ depends on four field parameters, we consider two schemes for systematically exploring the interplay of $a_s$ and $a_d$. 

In scheme $\text{I}$, $\Omega_\pi$ is varied with $\delta_\sigma, \Delta_\pi$ held fixed to first find the compensation point where the $\pi$ field exactly balances the $\sigma_+$ field, and the dipolar length $a_d$ goes to zero. Fig.~\ref{fig:1}(a) shows variation of the scattering length $a_s$ and the dipolar length $a_d$ with $\Omega_\pi$ for field parameters $(\Omega_\sigma,\Delta_\sigma,\Delta_\pi)=2\pi\times(8,8,10)$~MHz. 
Starting from the compensation point, $\Omega_\pi$ is varied towards the $\sigma_+$ dominated side close to the first field-linked resonance such that $a_s/|a_d|$ can be tuned to a large value. While scheme $\text{I}$ is good for experiments, it does not allow $a_s$ and $a_d$ to be varied independently of each other. Scheme $\text{II}$ overcomes this by varying any one of the field parameters while fixing $\delta_\sigma,\delta_\pi$ and the ratio of the two Rabi frequencies $\mathcal{G}=\Omega_\sigma/\Omega_\pi$. This fixes $a_d$ while the ratio $a_s/|a_d|$ can be tuned as a function of the chosen parameter.
Choosing $\Omega_\pi$ as the independent variable, $a_s/|a_d|$ becomes a universal function of the reduced Rabi frequency $\hbar\Omega_\pi/E_d$, independent of the molecular species for fixed $\delta_\sigma$, $\delta_\pi$, and $\mathcal{G}$. This reflects the fact that, in dipolar units, the microwave-dressed two-body Hamiltonian has the same dimensionless form at fixed reduced field parameters. 

\begin{figure*}[htbp]
    \includegraphics[width=\textwidth]{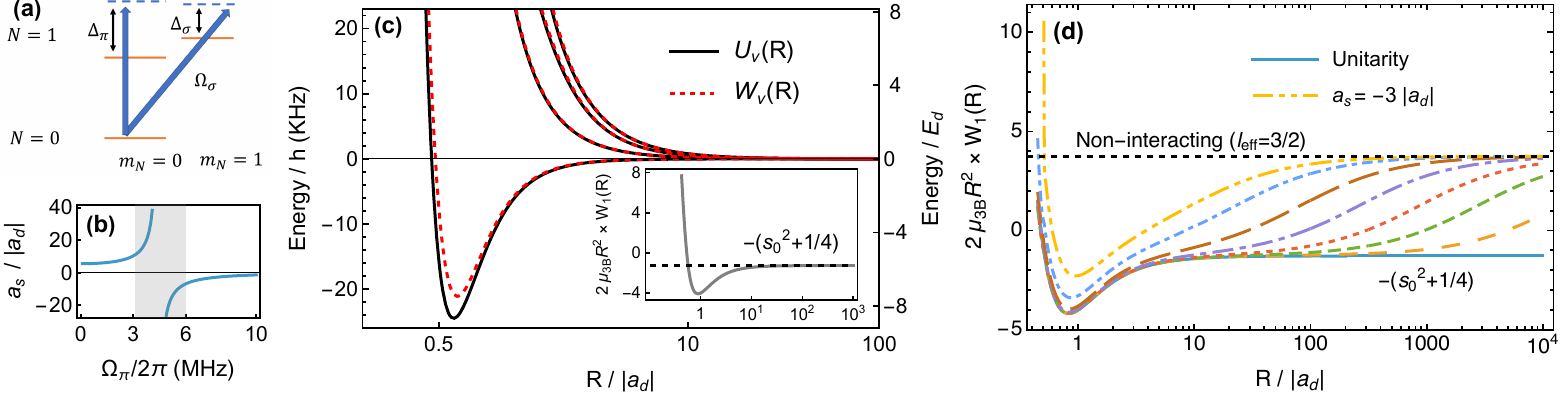}\\
    \caption{(a) Detuning scheme for double microwave shielding. (b)Variation of the dimensionless ratio $a_s/|a_d|$ in the vicinity of the first field linked resonance for CaF at field parameters $(\Omega_\sigma,\Delta_\sigma,\Delta_\pi)=2\pi\times(20,15,18)$ MHz. Grey shaded area shows the region where scattering length becomes the largest length scale and signatures of Efimov Physics start appearing. (c) Calculated three-body adiabatic potential curves for shielded CaF at unitarity $|a_s|\rightarrow\infty$. Adiabatic hyperspherical curves $U_{\nu}(R)$ (black solid) and $W_{\nu}(R)$ (red dashed) are shown together. Inset: Signature of the universal Efimov potential (d) Lowest hyperradial curves for different scattering lengths rescaled to reveal the effect of increasing $a_s/|a_d|$. Starting from the dashed-dot-dot curve, $a_s/|a_d|=-(3, ~7, ~26, ~83, ~273, ~864, ~4418)$ and the last solid one corresponds approximately to unitarity. Horizontal dashed line shows the non-interacting eigenvalue of the adiabatic Hamiltonian corresponding to $l_\text{eff}=3/2$. }
    \label{fig:2}
\end{figure*} 
Armed with a proper way to explore the parameter space of $V_\text{eff}$, the three-body potential is approximated as a pairwise sum of these two-body effective potentials, $V_{\text{{tot}}}=V_{\text{eff}}(\mathbf{r_{12}})+V_{\text{eff}}(\mathbf{r_{23}})+V_{\text{eff}}(\mathbf{r_{31}})$, the adiabatic hyperspherical description transforms the problem into coupled hyperradial equations written in atomic units as:
\begin{multline}
\left[
-\frac{1}{2\mu_\text{3B}} \frac{d^2}{dR^2} + U_\nu(R) - E
\right] F_\nu(R) \\
- \frac{1}{2\mu_\text{3B}} \sum_{\nu'} \left[
 2P_{\nu \nu'}(R)\frac{d}{dR} + Q_{\nu \nu'}(R)
\right] F_{\nu'}(R) = 0 .
\label{eq:1}
\end{multline}
where the hyperradius $R$ defined in the usual way \cite{greene2017universal} describes the size of the system, $\mu=m/\sqrt{3}$ is the three-body reduced mass, $\nu$ is a collective index for all the quantum numbers describing the channel, $E$ is the total energy, and $F_{\nu}$ and $U_{\nu}$ are respectively the hyperradial wavefunction and the adiabatic potential for the $\nu^{th}$ channel. $P_{\nu \nu'}(R)$ and $Q_{\nu \nu'}(R)$ are coupling matrices that drive non-adiabatic transitions. It is also useful to define the diagonally corrected adiabatic potential $W_\nu(R)=U_\nu(R)-Q_{\nu\nu}(R)/2\mu_{3B}$, whose lowest energy eigenvalue provides an upper bound to the lowest eigen energy in a given symmetry~\cite{upper_lowerbound}. The most challenging part of the problem is computing the adiabatic hyperspherical potential curves $U_\nu(R)$ by diagonalizing the adiabatic Hamiltonian $\hat{H}_\text{ad}\Phi_\nu(R; \vec\omega)=U_{\nu}(R)\Phi_\nu(R; \vec\omega)$, given in atomic units by:
\begin{equation}
\left[
\frac{{\hat{\Lambda}}^2(\vec\omega) + 15/4}{2\mu R^2} 
+ {V_{\text{tot}}}(R, \vec\omega) - U_\nu(R)
\right] \Phi_\nu(R; \vec\omega) = 0,
\label{eq:3}
\end{equation}
where $\Phi_{\nu}(R;\vec{\omega})$ is the channel eigenfunction, $\vec{\omega}$ represents the set of hyperangles and $\hat{\Lambda^2}(\vec{\omega})$ is the squared grand angular momentum operator with eigenvalues $\nu~(\nu+4)$. In a non-interacting system, eigenvalues of $\hat{\Lambda}^2(\vec{\omega})+15/4$ can be related to an effective angular momentum quantum number $l_{eff}=\nu+3/2$~\cite{delves1958tertiary}. The anisotropic nature of the dipole-dipole interaction in the presence of an external field means that the total three-body angular momentum $J$ is not a good quantum number, which further couples different $J$ blocks in the Hamiltonian. The channel functions are expanded in the body-fixed frame of the three-dipoles as:
\begin{equation}
 \Phi_\nu^{\Pi,M}(R; \vec\omega)=\sum_{J,K}\tilde{D}^J_{K,M}(\alpha,\beta,\gamma)\phi^{\Pi,J}_{K,M}(R;\theta,\varphi)
\end{equation}
Here, $\tilde{D}^J_{K,M}$ are appropriately normalized Wigner-D functions, $\tilde{D}^J_{K,M}=\sqrt{(2J+1)/8\pi^2}\,D^J_{K,M}$, defined in the passive sense~\cite{pack1987quantum,rose1995elementary} and $\alpha,\beta,\gamma$ are the Euler angles connecting the space fixed frame to the body fixed frame such that the plane containing the three bodies has no $z$-component in the body fixed frame, $\Pi$ is the parity, $J$ is the total angular momentum, $M$ is its projection on the space-fixed frame, and $K$ is its projection on the body-fixed frame. The functions $\phi^{\Pi,J}_{K,M}$ represent channel functions depending on the internal angles, while $\ 0\le \theta \le \pi/2,~0 \le \varphi \le \pi/3$ are the Smith-Whitten hyperangles, defined following the conventions of Ref.~\cite{chen2024universal}. The body-frame components $\phi^{\Pi,J}_{K,M}$ are obtained numerically by expanding in a two-dimensional B-spline basis in the variables $\theta$ and $\varphi$. Since this work is concerned with the Efimov effect, we restrict ourselves to $M^\Pi=0^+$, which corresponds to the $J^\Pi=0^+$ symmetry sector where the Efimov effect is known to appear even for dipolar systems~\cite{oi2024universality,wang2011efimov}, and include $J=0^+, 2^+$ and $4^+$ in the calculations. Furthermore, it is advantageous to work with linear combinations of $\tilde{D}^J_{K,M}$ given by $\tilde{D}^{J\pm}_{K,M}$. These combinations not only simplify the implementation of boundary conditions~\cite{SI} but also make the matrix elements purely real~\cite{SI,fano1960real}.

\begin{figure*}[t]
    \includegraphics[width=\textwidth]{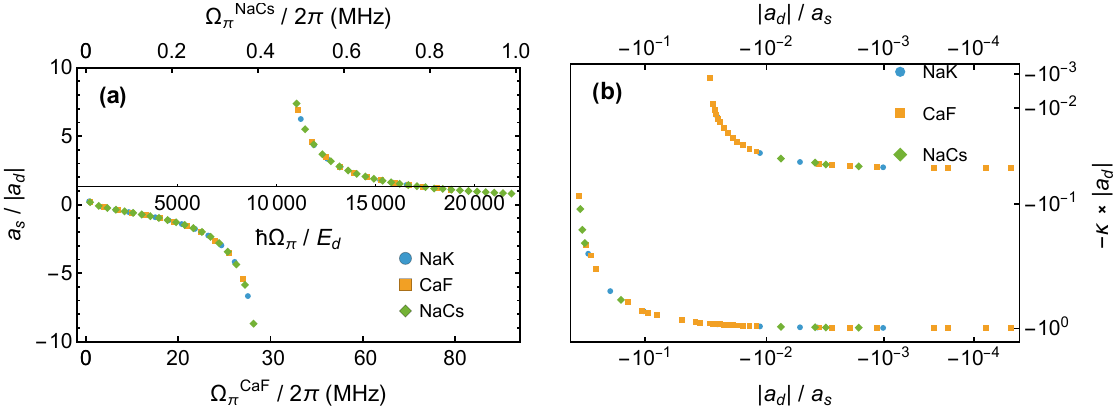}\\
    \caption{Two- and three-body universality. (a) $a_s/|a_d|$ for NaK (blue circles), CaF (orange squares) and NaCs (green diamonds) as a function of the reduced Rabi frequency $\hbar\Omega_\pi/E_d$ (middle x-axis), and the physical Rabi frequency for NaCs (top x-axis) and CaF (bottom x-axis) under scheme $\text{II}$ with $\delta_{\sigma,\pi}=(0.5,~2)$ and $\mathcal{G}=2$. For these field parameters, $a_d$ for NaCs is $-22795.4~a_0$, $a_d$ for NaK is $-3167.8~a_0$ while $a_d$ for CaF is $-3840.6~a_0$. (b) Three-body trimer spectrum for NaK (blue circles), CaF (orange squares) and NaCs (green diamonds) versus the inverse scattering length $|a_d|/a_s$
    }
    \label{fig:3}
\end{figure*}

 For field parameters corresponding to specific values of $a_s/|a_d|<-1$, the three-body adiabatic Hamiltonian is diagonalized for various effective two-body potentials. Fig.~\ref{fig:2}(b) shows how the scattering length $a_s$ varies in the vicinity of the first field linked resonance, and signatures of Efimov physics can be seen in the shaded area where $|a_s|>|a_d|$. Fig.~\ref{fig:2}(c) depicts the four lowest adiabatic hyperspherical potential curves $U_\nu(R)$ and $W_\nu(R)$ for $|a_s|\rightarrow \infty$ and the inset shows the lowest curve approaching the expected universal Efimov potential $-(s_0^2+1/4)/2\mu_{3B}R^2$ for $R\gg |a_d|$, providing a clear signature of Efimov effect existing in this system. Figure~\ref{fig:2}(d) shows how the lowest hyperspherical potential evolves with increasing $|a_s|/|a_d|$ toward the unitarity limit. Remarkably, the lowest three-body entrance channel always features a barrier at roughly $R\approx0.5~|a_d|$, which separates the three-body Efimov spectrum from the details of the short range. This is consistent with approximating the two-body potentials as a hard wall along with an attractive $1/r^4$ tail arising from a second order perturbation theory treatment with the dipolar and centrifugal terms~\cite{karman2025double}. Once $a_d$ is fixed, the shielding core sets $a_s$, whereas the hyperradial barrier near $R\sim0.5|a_d|$ arises from a reduced probability for three particles to be together due to increasing kinetic energy because of the attractive tail, similar to van der Waals universality~\cite{wang2012origin}. 
 In our studies, the universal Efimov potential emerges for all the molecules considered in this study: CaF, NaK and NaCs, provided that $|a_s|/|a_d|\gg1$. When expressed in dipolar units, the three-body spectrum of the molecular trimers with different field parameters collapse onto a universal curve. The interspecies collapse reflects two-body universality, whereas the geometric trimer spectrum and the resonance positions $a_n^-$ reflect Efimov universality. Figure~\ref{fig:3} summarizes this distinction, showing the collapse of $a_s/|a_d|$ in panel (a) and the interspecies collapse and scaling of trimer spectrum in panel (b).
\begin{table}[!htbp]
\centering
\caption{Three-body binding energies at unitarity and Efimov resonance positions in dipolar units,
shown with and without the diagonal correction.}
\begin{tabular}{lcccccc}
\hline\hline
 & $\kappa_0^{\infty} |a_d|$ & $\kappa_0^{\infty}/\kappa_1^{\infty}$ & $\kappa_1^{\infty}/\kappa_2^{\infty}$ & $a_0^-/|a_d|$ & $a_1^-/a_0^-$\\
\hline
No diag.\ corr.\ &-1.130  &18.52  &22.63  &-2.375 & 12.03  \\
With diag.\ corr.\ &-0.987  &20.14  &22.58  &-2.882 & 11.57 \\
\hline\hline
\end{tabular}
\end{table}

Table~I includes the Efimov resonance positions $a_n^-$, where an Efimov trimer crosses the three-body continuum, together with the three lowest trimer energies at unitarity in terms of $\kappa_n$ and the ratios $\kappa_n/\kappa_{n+1}$, both with and without the diagonal correction.
 The quantity $\kappa_n$ is defined as $\kappa_n=(m|E_n|)^{1/2}$, where $E_n$ is the trimer energy. Defining $\kappa_n^{\infty}$ as the $\kappa_n$ at unitarity, inclusion of the diagonal correction yields $\kappa_0^{\infty}/\kappa_1^{\infty}\approx20.14$, which deviates from the expected ratio of $e^{\pi/s_0}\approx22.7$, and this is not surprising since the wavefunction of lowest state in the spectrum has significant support in the region where the hyperradius $R\sim a_d$, but the outer turning point at unitarity still lies in the Efimov potential. The larger deviation of the ratio $a_1^-/a_0^-$ from the Efimov scaling factor, compared with the ratios $\kappa_n^\infty/\kappa_{n+1}^\infty$, indicates that the lowest trimer has more Efimov character at unitarity than at the threshold where it first becomes bound. The value of $\kappa_0^{\infty}=-0.987/|a_d|$ therefore defines a universal three-body parameter in dipolar units, which can be tuned through microwave field parameters via $a_d$. It should also be noted that usually the shape resonance at $a_n^-$ can act as an intermediate state that enhances the three-body recombination into deeper two-body states; however, for double microwave shielding, there are no shallow tetramers to recombine into, and the presence of the barrier at $R<a_d$ should further enhance the lifetime of these trimers, making this a favorable system for observing these molecular trimers.

\begin{figure}[htbp]
    \includegraphics[width=8.6cm]{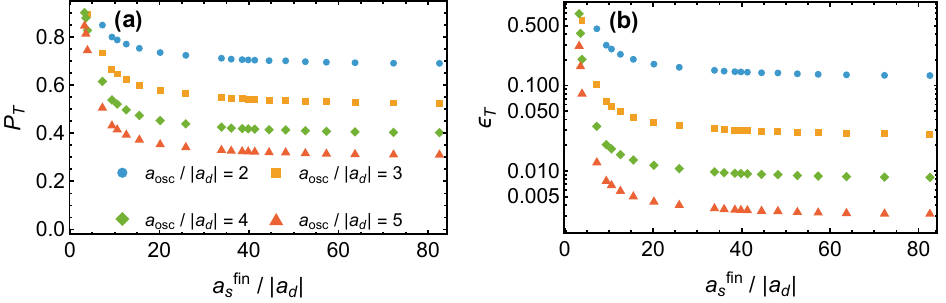}\\
    \caption{ (a) Calculated probability of trimer formation $P_T$ with $a_s^\text{in}=-2.5~|a_d|$ as a function of the final scattering length $a_s^\text{fin}/|a_d|$ for different values of the oscillator length $a_\text{osc}/|a_d|$. (b) Comparison of the absolute value of the calculated fractional difference $\epsilon_T$ as a function of $a_s^\text{fin}/|a_d|$ for the same set of oscillator lengths used in the left panel: $a_\text{osc}/|a_d|=2$~(blue circles), 3 (yellow squares), 4 (green diamonds), 5 (red triangles).
    }
    \label{fig:4}
\end{figure}

Next, we estimate the feasibility of producing these trimers from a trapped molecular sample in an optical dipole trap. For an isotropic harmonic confinement with frequency $2\pi \times \omega_{\text{trap}}$, the hyperradial equation in center of mass frame is
\begin{multline}
    \left[-\frac{1}{2\mu_{\text{3B}}}\frac{\partial^2}{\partial R^2}+W_{\nu}^{a_s}(R)\right.\\\left.
    +\frac{1}{2}\mu_\text{3B}\omega_{trap}^2 R^2\right]F^{a_s}_{\nu,i}(R) =E_{\nu,i}F^{a_s}_{\nu,i}(R),
\end{multline}
where $a_{\text{osc}} = (\mu_{\text{3B}}\,\omega_{\text{trap}})^{-1/2}$, $W^{a_s}_{\nu}(R)$ is the three-body adiabatic potential, and $F^{a_s,a_{\text{osc}}}_{\nu,i}(R)$ is the hyperradial solution with energy $E^{a_s,a_{osc}}_{\nu,i}$. The initial scattering length $a_s^{in}$ and oscillator length $a_{osc}$ are chosen such that the initial Hamiltonian supports no free-space three-body bound state and the eigenstates are trap dominated, $E^{a_s,a_{osc}}_{0 i}>0 ~\forall ~i\geq0$, where $E=0$ corresponds to the three-body continuum. After a sudden quench to the final scattering length $a_s^{fin}$, the final Hamiltonian can support an Efimov trimer in the same lowest adiabatic channel. Focusing on this lowest adiabatic potential, the initial wavefunction is prepared in the lowest eigenstate of $W_0^{a_\text{in}}(R)+1/2\mu_{3B}\omega_\text{trap}^2R^2$ given by $\ket{\psi^\text{in}}=F^\text{in}_{0,0}(R)\ket{\Phi^\text{in}_0(R;\vec{\omega})}$. The final wavefunction for the target trimer state at scattering length $a_s^\text{fin}$ is $\ket{\psi^\text{fin}}=F^\text{fin}_{0,0}(R)\ket{\Phi^\text{fin}_0(R;\vec{\omega})}$, where the oscillator length is held fixed. In the sudden quench limit, the probability to form a trimer from an initially occupied $\nu=0$ trap eigenstate is $P_{T}=|\int F^\text{fin}_{0,0}(R)F^{\text{in}}_{0,0}(R)S(R)~dR|^2$, where the hyperangular overlap $S(R)=\braket{\Phi^{\text{fin}}_0(R;\vec\omega)}{\Phi^{\text{in}}_0(R;\vec\omega)}_{\vec\omega}$ is an integral over the hyperangles $\vec{\omega}$. Fig.~\ref{fig:4}(a) shows $P_T$ versus $a_s^{\text{fin}}$ for different oscillator lengths. Higher overlap for smaller value of $a_\text{osc}$ can be misleading as $\ket{\psi^\text{fin}}$ has both trap and Efimov character, meaning that the trap character can dominate the overlap. We estimate the Efimov character by comparing $E^\text{fin}_{0,0}(a_\text{osc})$ with the free-space trimer energy $E^\text{fin}_{0,0}(a_\text{osc}\rightarrow\infty) \equiv E^\text{fin}_T$. Fig.~\ref{fig:4}(b) considers the absolute value of this fractional difference  $\epsilon_T(a_{\text{fin}},a_\text{osc})=|(E^\text{fin}_T-E^\text{fin}_{0,0}(a_\text{osc}))/E^\text{fin}_T|$, showing that eigenstates at higher oscillator lengths, despite having a lower overlap with $\ket{\psi^\text{in}}$, have more Efimov character. Table~II gives representative NaK parameter sets in physical units for the quench protocol, along with the resulting overlap and $\epsilon_T$ values. Experimentally, one possible route to verify the associated trimers would be to weakly modulate a dressing field parameter to observe a resonant depletion signal in the molecule population, similar to work on NaK tetramers~\cite{chen2024ultracold}. Another possibility could be using Ramsey-type interferometry using modulated magnetic field pulses to obtain binding energies and lifetimes~\cite{bougas2023interferometry}.
\begin{table}[!htbp]
\centering
\caption{Representative parameter sets for the quench protocol in physical units for NaK with $a_d=-3167.8~a_0$. The reported overlap gives the projected population in the target trimer state of the final Hamiltonian.}
\begin{tabular}{ccccc}
\hline\hline
$\omega_{\mathrm{trap}}/2\pi$ (Hz) & $a_s^{\mathrm{in}}$ ($10^3 a_0$) & $a_s^{\mathrm{fin}}$ ($10^3 a_0$) & Overlap (\%) & $\epsilon_T$ \\
\hline
2505.14 & -7.92 & -107.39 & 71.2 & 0.16  \\
1113.39  & -7.92 & -63.99  & 57.9 & 0.03  \\
626.28  & -7.92 & -49.10  & 47.4 & 0.013 \\
400.82  & -7.92 & -49.10  & 37.8 & 0.005 \\
\hline\hline
\end{tabular}
\label{tab:assoc_params}
\end{table}

In conclusion, we have developed a quantum-mechanical description for the three-body physics of shielded dipolar molecules simultaneously dressed by $\sigma_+$ and $\pi$ fields, explicitly incorporating the full anisotropic interaction in three dimensions. Despite these complications, shielding opens up a regime where Efimov physics emerges. When expressed in dipolar units, both two- and three-body adiabatic potentials are determined entirely by a single dimensionless ratio $a_s/|a_d|$. The presence of a universal barrier in the lowest three-body potential at $R\sim0.5a_d$, isolates the Efimov spectrum from the short range. Consequently, the trimer energies and positions of Efimov resonances are completely determined by $a_s/|a_d|$, and the three-body parameter is universal in dipolar units, while remaining tunable in absolute units through control over the dipolar length $a_d$.  

The computed trimer energies exhibit the universal geometric scaling expected for Efimov states. Finally, a rapid quench of the scattering length from the trap dominated regime into the Efimov regime yields a non-zero overlap with the target trimer state, providing a qualitative mechanism for producing these trimers starting from a trap state. While this work focuses on bosonic molecules, future studies can consider fermionic species, or heteronuclear systems, or explorations of the influence of field ellipticity on the Efimov spectrum. Broadly, our results establish shielded polar molecules as a highly tunable platform to explore few-body physics in molecular systems. 

We thank Ian Stevenson and Sebastian Will for helpful discussions during the early stages. This work was supported in part by NSF Grant No. PHY-2512984.

\begin{widetext}
\end{widetext}




\clearpage
\onecolumngrid
\input{supplemental_body}
\clearpage
\twocolumngrid
\bibliography{bib2}             

\end{document}

%% file: supplemental_body.tex
\clearpage
\onecolumngrid

\setcounter{section}{0}
\setcounter{equation}{0}
\setcounter{table}{0}

\renewcommand{\thesection}{S\arabic{section}}
\renewcommand{\theequation}{S\arabic{equation}}
\renewcommand{\thetable}{S\arabic{table}}

\section*{Supplemental Material for ``Efimov Effect in Ultracold Microwave-Shielded Polar Molecules''}
\section{Effective two-body potentials}
The effective potential for the shielded channel~\cite{deng2025two} is rewritten for convenience as:
\begin{equation}
    V_{\text{eff}}(\mathbf{r})=-\frac{C_{3}}{r^3}\sqrt{\dfrac{16\pi}{5}}Y_{2,0}(\hat{\mathbf{r}})+\frac{C_6}{r^6}\sum_{q=0}^2 \tilde{w}_q|Y_{2,q}(\hat{\mathbf{r}})|^2,
\end{equation}
where the coefficients $\tilde{w}_0=16\pi/5~w_0,~\tilde{w}_1=8\pi/15~w_1$ and $\tilde{w_2}=32\pi/15~w_2$. The two-body adiabatic potentials are obtained by diagonalizing in a basis of partial waves $\ket{l,m}$, representing the relative angular momentum and it's projection on the space-fixed z-axis.  The matrix element for the effective potential is given by
\begin{align} 
\label{ad ham}
    \bra{l',m'}\hat{V}_{eff}(\vec{r})\ket{l,m} = -\frac{C_{3}}{r^3}\sqrt{\dfrac{16\pi}{5}}\mathcal{I}^{l',m'}_{2,0,l,m}+\frac{C_6}{r^6}\sum_{q,L}^2 (-1)^q \tilde{w}_q\mathcal{I}^{L,0}_{2,q,2,-q}\,\mathcal{I}^{l',m'}_{L,0,l,m},
\end{align}
where $0\leq L \leq 4$ is restricted by triangularity and the integral of three spherical harmonics is given by
\begin{equation}
\mathcal{I}^{l',m'}_{L,M,l,m} = \int Y_{l',m'}^{*}(\hat{\mathbf{r}})Y_{L,M}(\hat{\mathbf{r}})Y_{l,m}(\hat{\mathbf{r}})\,d\Omega
= \sqrt{\dfrac{(2L+1)(2l+1)}{4\pi\,(2l'+1)}} C^{l',0}_{L,0,l,0}C^{l',m'}_{L,M,l,m},\end{equation}
and $C^{LM}_{l,m,l'm'}$ are Clebsch-Gordan coefficients in the Condon-Shortley phase convention.
The two-body coupled equations are solved using the slow-variable discretization (SVD) method~\cite{tolstikhin1996slow} along with the discrete variable representation (DVR) to propagate the R-matrix~\cite{wang2012hyperspherical,manolopoulos1988quantum} from $10^2 a_0\leq r\leq10^7a_0$ with 1135 DVR sectors, including even partial waves up to $l=16$. The K-matrix and S-matrix are readily obtained from the R-matrix by imposing asymptotic boundary conditions at large $r$, using the energy-normalized spherical Bessel and Neumann functions:
\begin{subequations}\
\begin{equation}
    \mathbf{K}=(\mathbf{f}-\mathbf{f'R})(\mathbf{g-g'R})^\mathbf{-1}
\end{equation}
\begin{equation}
    \mathbf{S}=(\mathbf{I+}i\mathbf{K})(\mathbf{I-}i\mathbf{K})^\mathbf{-1},
\end{equation}
\end{subequations}
 and the diagonal matrices $\mathbf{f},\mathbf{g}$ are given in terms of the spherical Bessel $j_l(k_lr)$ and Neumann $n_l(k_lr)$ functions by:
\begin{subequations}
    \begin{equation}
        f_{l'm',lm}=\sqrt{\frac{2\mu_\text{2B}}{\pi k_{lm}}}k_{lm} r j_{l}(k_{lm} r)\delta_{ll'}\delta_{m'm}
    \end{equation}
    \begin{equation}
        g_{l'm',lm}=\sqrt{\frac{2\mu_\text{2B}}{\pi k_{lm}}}k_{lm} r n_{l}(k_{lm} r)\delta_{ll'}\delta_{m'm},
    \end{equation}
\end{subequations}
where $\mu_{2B}=m/2$, $k_{lm}=(2\mu_\text{2B}(E-E_{lm}))^{1/2}$ and $E_{lm}$ is the threshold energy in the asymptotically $\ket{l,m}$ dominated channel, the diagonal matrices $\mathbf{f'},\mathbf{g'}$ are just derivatives of $\mathbf{f,g}$. The K-matrix element $K_{l'm',lm}$ represents incoming wave in $\ket{l,m}$ channel and outgoing wave in $\ket{l',m'}$ channel. Scattering length for the S-wave dominated channel is defined as $a_s=\lim_{k_{00}\rightarrow0}-K_{00,00}/k_{00}$.

\section{Adiabatic Hyperspherical Representation}
In the center of mass frame, the three-body system can be represented by the mass scaled Jacobi coordinates as:
\begin{subequations}
\begin{equation}
    \vec\rho^{~(k)}_1 = \frac{\vec{r}_j-\vec{r}_i}{d_{ij}}
\end{equation}
\begin{equation}
    \vec{\rho}^{~(k)}_{2}=d_{ij}\left(\vec{r}_k-\frac{m_i\vec{r}_i+m_j\vec{r}_j}{m_i+m_j}\right)
\end{equation}
\end{subequations}
Here, $m_i$ and $\vec{r}_i$ denote the mass and position vector of molecule $i$, respectively; the superscript $(k)$ labels the odd particle in this particular Jacobi tree, and for three identical molecules one has $d_{ij}=2^{1/2}3^{1/4}$. The hyperradius $R$ is defined via $R^2 = |\vec{\rho}^{(k)}_i|^2 + |\vec{\rho}^{(k)}_j|^2$ and is independent of the particular Jacobi tree chosen in its construction. Within the adiabatic hyperspherical formalism, $R$ is treated as an adiabatic parameter. Having excluded the center of mass, of the remaining five degrees of freedom, three are eliminated using the Euler angles $\alpha, \beta, \gamma$ to rotate into the body-fixed frame of the three-particles, where the body-fixed z-axis is chosen to be parallel to $\vec{\rho}^{(k)}_i \times \vec{\rho}^{(k)}_j$. The internal motion is described by modified Smith-Whitten coordinates $\theta,\varphi$, defined by the components of the Jacobi vectors in the body-fixed frame~\cite{suno2008adiabatic} by
\begin{subequations}
\begin{equation}
 \vec{\rho}^{~(k)}_{1x}=R\cos{(\pi/4-\theta/2)}\cos{(\varphi/2+\varphi_{ij}/2)},   
\end{equation}
\begin{equation}
 \vec{\rho}^{~(k)}_{1y}=R\sin{(\pi/4-\theta/2)}\sin{(\varphi/2+\varphi_{ij}/2)},   
\end{equation}
\begin{equation}
   \vec{\rho}^{~(k)}_{2x}=-R\cos{(\pi/4-\theta/2)}\sin{(\varphi/2+\varphi_{ij}/2)}, 
\end{equation}
\begin{equation}
    \vec{\rho}^{~(k)}_{2y}=R\sin{(\pi/4-\theta/2)}\cos{(\varphi/2+\varphi_{ij}/2)},
\end{equation}
\end{subequations}
and $\vec{\rho}^{~(k)}_{1z}=\vec{\rho}^{(k)}_{2z}=0$. For a system of identical particles, one has $\varphi_{12}=2\pi/3$, $\varphi_{23}=0$, $\varphi_{31}=-2\pi/3$, with the hyperangles restricted to the intervals $\varphi\in[0,\pi/3]$ and $\theta\in[0,\pi/2]$. Using the Smith-Whitten hyperangles, the three-body Schrodinger equation can be written as~\cite{johnson1983quantum}
\begin{equation}
    \left\{-\frac{1}{2\mu_{3B}}\left[\frac{1}{R^5}\frac{\partial}{\partial R}R^5\frac{\partial}{\partial R}-\frac{\hat{\Lambda}^2}{R^2}\right]+\hat{V}_{\text{tot}}(R,\vec{\omega})\right\}\Psi = E\Psi
\end{equation}
where, $\hat{\Lambda}^2$ operator is given by
\begin{equation}
\hat{\Lambda}^2 =  -\frac{4}{\sin{2\theta}}\frac{\partial}{\partial\theta}\sin{2\theta}\frac{\partial}{\partial \theta}+\frac{4}{\sin^2{\theta}}\left(i\frac{\partial}{\partial\varphi}-\cos{\theta}\frac{\hat{J}_{z'}}{2}\right)^2 +\frac{2\hat{J}_{x'}^2}{1-\sin{\theta}}
+\frac{2\hat{J}_{y'}^2}{1+\sin{\theta}}+\hat{J}_{z'}^2.
\end{equation}
Here, $\hat{J}_{x'},\hat{J}_{y'},\hat{J}_{z'}$ are body frame cartesian components of the total angular momentum operator, which satisfy the anomalous commutation relations $[\hat{J}_{i'},\hat{J}_{j'}]=-i\epsilon_{ijk}\hat{J}_{k'}$. Definitions of $\hat{J}_{x'},\hat{J}_{x'},\hat{J}_{y'}$ in terms of the Euler angles can be found in Ref.~\cite{johnson1983quantum}.

\section{Boundary condition for channel functions}
It is instructive to derive the boundary conditions obeyed by the channel functions in the internal hyperangles $(\theta,\varphi)$. For a given $J^\Pi$ and $M=0$, the channel function is expanded as 
\begin{equation}
    \Phi_\nu^{\Pi,J,0}(R; \vec{\omega}) = \sum_K \phi_{\nu,K}^{\Pi,J}(R; \theta,\varphi) \tilde{D}^J_{K,0}(\alpha,\beta,\gamma),
    \label{eq:14}
\end{equation}
where $\hat{\Pi}$ only affects $\tilde{D}^J_{K,M}$ such that $\hat{\Pi}\Phi_\nu^{\Pi,J,M}=(-1)^K\Phi_\nu^{\Pi,J,M}$, restricting $K$ to odd or even depending on the parity, $M=0$ in the superscript is suppressed for $\phi(R;\theta,\varphi)$, the sum over $K$ runs from $-J,~-(J-2),~\dots,~J-2,~
J$ for parity favored states and from $-(J-1),~-(J-3),~\dots,~J-3,~J-1$ for parity unfavored states. Another useful identity is the continuity condition for the channel functions given by
\begin{equation}
    \Phi_\nu^{\Pi,J,M}(R; \theta,\varphi,\alpha,\beta,\gamma) = \Phi_\nu^{\Pi,J,M}(R; \theta,\varphi+2\pi,\alpha,\beta,\gamma+\pi)
\end{equation}
Grouping terms with the same absolute value of $K$ and using $d^J_{K,M}(\beta) = (-1)^{M-K}d^J_{-K,-M}(\beta)$, Eq.~\ref{eq:14} can be rewritten as
\begin{equation}
    \sum_K \phi^{\Pi,J,+}_{\nu,K} \tilde{D}^{J,+}_{K,0}+\phi^{\Pi,J,-}_{\nu,K}\tilde{D}^{J,-}_{K,0}
\end{equation}
where $\tilde{D}^{J,\pm}_{K,M}$ are linear combinations of  Wigner D-functions given by:
    \begin{equation}
        \tilde{D}^{J,\pm}_{K,0}=\dfrac{\tilde{D}^{J}_{K,0}+(-1)^{K+\eta}\tilde{D}^{J}_{-K,0}}{i^{\eta}\sqrt{2~(1+\delta_{K,-K})}}
    \end{equation}
where $\tilde{D}^{J,+}_{K,M}$ and $\tilde{D}^{J,-}_{K,M}$ correspond to $\eta=0$ and $\eta=1$ respectively. Table~\ref{tab:perm_phi_wignerD} summarizes the action of the permutation operators on the the internal hyperangle $\varphi$ and Wigner-D functions $D^J_{K,M}$. Now under the action of the permutation operator $\hat{P}_{23}$, using Table ~\ref{tab:perm_phi_wignerD}, $\hat{P}_{23}D^J_{K,M}=(-1)^JD^J_{-K,M}$. One can show that applying any $\hat{P}_{ij}$ on the linear combinations $\tilde{D}^{J,\pm}_{K,M}$ yields $\hat{P}_{ij}\tilde{D}^{J\pm}_{KM}=(-1)^{J+K+\eta}\tilde{D}^{J\pm}_{K,M}$, making these linear combinations a very convenient choice for applying boundary conditions obeying the permutation symmetry. 
\begin{table}[t]
\caption{Effects of permutation operations on the hyperangle $\varphi$ and the Wigner $D$ function.}
\label{tab:perm_phi_wignerD}
\begin{ruledtabular}
\begin{tabular}{ccc} 
Permutation & $\varphi$ & $D^{J}_{KM}$\\
\hline
\\
$P_{12}$      & $\dfrac{2\pi}{3}-\varphi$ & $(-1)^{J} D^{J}_{-K\,M}$ \\ \\
$P_{23}$      & $\dfrac{4\pi}{3}-\varphi$ & $(-1)^{J}   D^{J}_{-K\,M}$ \\ \\
$P_{31}$      & $2\pi-\varphi$            & $(-1)^{J}   D^{J}_{-K\,M}$ \\ \\
$P_{12}P_{31}$& $\varphi+\dfrac{2\pi}{3}$ &    $D^{J}_{K\,M}$ \\ \\
$P_{12}P_{23}$& $\varphi-\dfrac{4\pi}{3}$ & $            D^{J}_{K\,M}$ \\ \\
\end{tabular}
\end{ruledtabular}
\end{table}
For identical bosons, applying $\hat{P}_{12}$ operator on the channel function should leave it invariant: 
\begin{subequations}
\begin{equation}
    \hat{P}_{12}\Phi_\nu^{\Pi,J,0}(R; \vec{\omega})=\Phi_\nu^{\Pi,J,0}(R; \vec{\omega})=\sum_K \phi^{\Pi,J,+}_{\nu,K}(
2\pi/3-\varphi) (-1)^{J+K}\tilde{D}^{J,+}_{-K,0}-\phi^{\Pi,J,-}_{\nu,K}(2\pi/3-\varphi)(-1)^{J+K}\tilde{D}^{J,-}_{-K,0}
\end{equation}
Focusing on the parity-favored states with even $J$,
\begin{equation}
    \phi^{\Pi,J,+}_{\nu,K}(2\pi/3-\varphi)=\phi^{\Pi,J,+}_{\nu,K}(\varphi)\implies \dfrac{\partial}{\partial \varphi}\phi^{\Pi,J,+}_{\nu,K}(\varphi=\pi/3) = 0 
\end{equation}
\begin{equation}
    \phi^{\Pi,J,-}_{\nu,K}(2\pi/3-\varphi)=-\phi^{\Pi,J,-}_{\nu,K}(\varphi)\implies \phi^{\Pi,J,-}_{\nu,K}(\varphi=\pi/3) = 0
\end{equation}
\end{subequations}
Similarly, applying the $\hat{P}_{31}$ operator on identical bosons, and using with the continuity condition yields the following boundary conditions for parity-favored states with even $J$:
\begin{subequations}
\begin{equation}
    \phi^{\Pi,J,+}_{\nu,K}(-\varphi)=\phi^{\Pi,J,+}_{\nu,K}(\varphi)\implies \dfrac{\partial}{\partial \varphi}\phi^{\Pi,J,+}_{\nu,K}(\varphi=0) = 0 
\end{equation}
    \begin{equation}
          \phi^{\Pi,J,-}_{\nu,K}(-\varphi)=-\phi^{\Pi,J,-}_{\nu,K}(\varphi)\implies \phi^{\Pi,J,-}_{\nu,K}(\varphi=0) = 0
    \end{equation}
\end{subequations}
In general, the boundary conditions can be summarized using the following equations:
\begin{subequations}
\begin{equation}
    \phi^{\Pi,J,\pm}_{\nu,K}(2\pi/3-\varphi)=(-1)^{J+K+\eta}\phi^{\Pi,J,\pm}_{\nu,K}(\varphi)
\end{equation}
    \begin{equation}
          \phi^{\Pi,J,\pm}_{\nu,K}(-\varphi)=(-1)^{J+\eta}\phi^{\Pi,J,\pm}_{\nu,K}(\varphi)
    \end{equation}
\end{subequations}
where again $\eta=0,1$ corresponds to $\tilde{D}^{J,\pm}_{K,M}$ respectively.
\section{Matrix Elements of the Hamiltonian}
 During the calculation of matrix elements, it is convenient to rewrite $\hat{\Lambda}^2$ as
\begin{subequations}
\begin{equation}
    \hat{T}_1 = -\frac{4}{\sin{2\theta}}\frac{\partial}{\partial\theta}\sin{2\theta}\frac{\partial}{\partial \theta}-\frac{4}{\sin^2{\theta}}\frac{\partial}{\partial \varphi^2}+\frac{2(\hat{J}_{x'}^2+\hat{J}_{y'}^2)}{\cos^2{\theta}}+\frac{\hat{J}_{z'}^2}{\sin^2{\theta}}
\end{equation}
\begin{equation}
    \hat{T}_2 = \frac{\sin{\theta}(\hat{J}_+^2+\hat{J}_-^2)}{\cos^2{\theta}}+\frac{4i\cos{\theta}}{\sin^2{\theta}}\hat{J}_{z'}\frac{\partial}{\partial \varphi}
\end{equation}
\end{subequations}
It is straightforward to verify that $\hat{T}_1$ is diagonal in the basis $\tilde{D}^{J\pm}_{K0}\equiv\ket{J,K,\pm}$. The first term in $\hat{T}_2$ connects $\ket{J,K,\pm}$ to $\ket{J,K\pm2,\pm}$ but does not couple $\ket{J,K,+}$ to any $\ket{J,K',-}$. More significantly, the second term links $\ket{K,+}$ and $\ket{K,-}$, and in this representation $\hat{J}_{z'}$ is purely imaginary, which ensures that the matrix elements of the $i\hat{J}_{z'}$ term are real.

Evaluating the matrix elements of the interaction potential involves terms of the form:
\begin{equation}
    \left\langle 
\phi^{\Pi,J',\pm}_{\nu',K'} , J',K',\pm
\right|
\hat{V}_{\mathrm{eff}}(\vec r_{ij})
\left|
J,K,\pm , \phi^{\Pi,J,\pm}_{\nu,K}
\right\rangle.
\end{equation}
This is done in two steps. The first step amounts to doing the integral over the Euler angles analytically. Note that $V_\text{eff}(\vec{r}_{ij}(R,\vec{\omega}))$ contains spherical harmonics $Y_{lm}(\hat{r}_{ij})$ defined with respect to the space fixed $z$-axis. Expanding in terms of spherical harmonics in the body-fixed frame:
\begin{equation}
    Y_{l,m}(\hat{r}_{ij})=\sum_{\Lambda}D^l_{\Lambda,m}(\alpha,\beta,\gamma)Y_{l,\Lambda}(\hat{r'}_{ij})
    \label{eq:19}
\end{equation}
where, $\hat{r'}_{ij}$ is defined in the body-fixed frame and $\Lambda$ is the projection on the body-fixed $z'$ axis and using the fact that:
\begin{equation}
    \tilde{D}^J_{K0}(\alpha,\beta,\gamma)=\sqrt{\frac{2J+1}{8\pi^2}}D^J_{K,0}(\alpha,\beta,\gamma)=\frac{(-1)^K}{\sqrt{2\pi}}Y_{J,K}(\beta,\gamma),
    \label{eq:20}
\end{equation}
the matrix elements over the Euler angles reduce to integral over three-spherical harmonics, which can be done analytically. The second step involves a numerical integration over the internal hyperangles $(\theta,\varphi)$ using quadratures.